\documentclass[12pt]{iopart}

\usepackage{graphicx}
\begin{document}

\title{First demonstration of multi-MeV proton acceleration from a cryogenic hydrogen ribbon target}

\author{Stephan D. Kraft$^1$, Lieselotte Obst$^{1,2}$ , Josefine Metzkes-Ng$^1$, Hans-Peter Schlenvoigt$^1$, Karl Zeil$^1$, Sylvain Michaux$^3$, Denis Chatain$^3$, Jean-Paul Perin$^3$, Sophia N. Chen$^4$, Julien Fuchs$^4$, Maxence Gauthier$^5$, Thomas E. Cowan$^{1,2}$, Ulrich Schramm$^{1,2}$}
\address{$^1$Institute of Radiation Physics, Helmholtz-Zentrum Dresden-Rossendorf, Dresden, Germany

$^2$Technische Universit\"at Dresden, Dresden, Germany

$^3$University of Grenoble Alpes, CEA INAC-SBT, Grenoble, France

$^4$LULI - CNRS, Ecole Polytechnique, CEA: Université Paris-Saclay; UPMC Univ Paris 06: Sorbonne Universités, F-91128, Palaiseau cedex, France

$^5$High Energy Density Science Division, SLAC National Accelerator Laboratory, Menlo Park, California, 94025, USA
}

\ead{s.kraft@hzdr.de}

\begin{abstract}
We show efficient laser driven proton acceleration up to 14\,MeV from a 50\,$\mu$m thick cryogenic hydrogen ribbon. Pulses of the short pulse laser ELFIE at LULI with a pulse length of $\approx 350$\,fs at an energy of 8\,J per pulse are directed onto the target. The results are compared to proton spectra from metal and plastic foils with different thicknesses and show a similar good performance both in maximum energy as well as in proton number. Thus, this target type is a promising candidate for experiments with high repetition rate laser systems.


\end{abstract}

\maketitle

\section{\label{Intro}Introduction}
 
Over the last decades, ion beams accelerated with ultrahigh power lasers have become more and more reliable \cite{Macchi2013}. In addition, the maximum proton energies have increased and it is now possible to accelerate protons to maximum energies of 85 MeV \cite{Wagner2016} with the Target Normal Sheath Acceleration mechanism \cite{Snavely2000}. Compared to ion beams accelerated with conventional accelerators, laser plasma accelerated bunches exhibit an exponential energy spectrum with a well-defined maximum cut-off energy and unprecedented particle numbers of up to $10^{12}$ per shot. The ion bunches have a short bunch duration of only a few picoseconds at the source and can thus be used to apply a very high dose rate. These characteristic properties make laser driven ion beams interesting for many applications \cite{Ledingham2010}. Among them, their application to proton beam therapy to treat cancer is surely one of the most prominent \cite{Malka2004,Yogo2009,Kraft2010,Masood2017}.

In parallel to the advances in ion acceleration, we face a rapid development in laser technology. In recent years, several lasers with powers exceeding one Petawatt came into operation worldwide \cite{Danson2015,Schramm2017}. Some of these lasers have repetition rates in the range of one shot every second making these facilities promising candidates for aforementioned applications. This development leads to a need for new target types as ion sources that are suitable for high repetition rate  operation \cite{Prencipe2017}: First, target assemblies have to allow for a high number of shots before exchange of the target material becomes necessary, which usually requires interrupting the ion acceleration run. Secondly, mitigating the production of debris,  which results from the vaporization of the target during laser interaction, increases the life-time of expensive optics and is therefore crucial when going to high shot-rates.
In order to meet these requirements, several different approaches are possible. The target can be made very thin in order to have less material interacting with the laser \cite{Poole2014,Seuferling2017}. A second possibility is the use of dense gas jets for proton acceleration, which do not produce any debris at all \cite{Semushin2001,Willingale2006,Palmer2011,Chen2017}.

In addition to the mentioned possibilities, cryogenic hydrogen targets are promising candidates for proton acceleration. Due to the absence of other atomic species, intrinsically no debris is produced. Compared to the experiments with gas jets, cryogenic hydrogen is overcritical which enables the robust Target Normal Sheath Acceleration mechanism for laser-driven ion acceleration \cite{Snavely2000}. Up to now several different approaches have been implemented. 

The first possibility is a free standing hydrogen foil \cite{Astbury2016,Tebartz2017}. The foils are produced in a frame prior to each shot in the vacuum chamber. As one laser pulse destroys an entire hydrogen foil, they have to be replaced before every single shot. A further possibility is a cryogenic hydrogen jet \cite{Gauthier2016,Obst2017}. Here, a liquid hydrogen microjet enters the target chamber through an aperture to produce the desired target geometry and freezes out. This target is produced continuously and enables shots with high repetition rates \cite{Gauthier2017}. 

The third implementation is a cryogenic ribbon extruded from a reservoir of solid hydrogen \cite{Garcia2014}. In contrary to the free standing targets, the ribbon is renewable and thus can be used for multiple shots. Since the amount of hydrogen injected into the chamber is low, no additional vacuum pumps have to be installed to the chamber. The ribbon target was tested successfully at a long pulse laser system with 0.3\,ns pulse duration and an energy of 600\,J per laser pulse \cite{Margarone2016}. Although these laser types are not normally used for particle acceleration, protons were accelerated to energies in the range of 1\,MeV.

In this paper, we present the first proton acceleration results from  solid hydrogen ribbons using a short pulse laser showing a similarly performance in proton acceleration in comparison with metal and plastic foils.

\section{Experimental setup}

The presented results were collected in an experiment carried out at  the ultra high intensity laser system ELFIE at LULI, France. The laser was operated at a wavelength of 1053\,nm. The laser pulses had a pulse length of $\approx 350$\,fs at an energy of 8\,J per pulse. The pulses were focused with a parabolic mirror to a spot size of 6\,$\mu$m x 4.5\,$\mu$m full width half maximum resulting in an intensity of
$10^{19}$\,W/cm$^2$. 
The  normalized temporal intensity contrast of the laser was $10^{-7}$ up to 1.5\,ps before the maximum intensity. 

\begin{figure}
\centering
\includegraphics[width=.9 \textwidth]{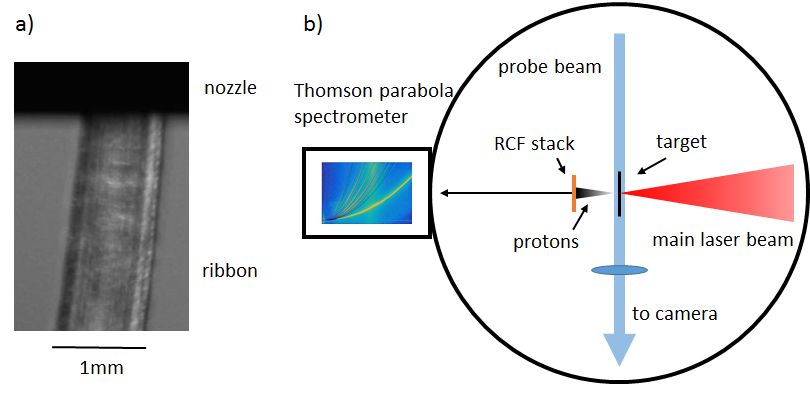}
\caption{a) Front view of the hydrogen ribbon emerging from the cold nozzle: The ribbon has a width of roughly 1\,mm and a thickness of either 50\,$\mu$m or 100\,$\mu$m, depending of the nozzle used in the setup. The length of the ribbon can be up to several centimeters. b) Sketch of the top-view of the ELFIE target chamber. The laser is focused tightly onto the target. Protons are detected by a Thomson parabola and RCF stacks.}
\label{setup}
\end{figure}

The hydrogen ribbon was produced with the ELISE (experiments on laser interaction with solid hydrogen) cryostat developed at CEA Grenoble, France \cite{Garcia2014}. The cryostat produces a solid ribbon from a reservoir of frozen hydrogen. The main component of ELISE contains a reservoir for the hydrogen. An exchangeable nozzle can be attached to the lower part of this reservoir. After filling liquid hydrogen into the reservoir it is cooled to a temperature near the triple point of hydrogen ($\approx$ 12\,K) and the hydrogen freezes out. By heating the upper part of the reservoir, the now solid hydrogen is pressed through the nozzle with a pressure exceeding 100\,bar. The hydrogen forms a 1\,mm wide ribbon which flows continuously into the vacuum chamber (see fig.\,\ref{setup}a)). The thickness of the ribbon is given by the dimensions of the nozzle and can be set to either 50\,$\mu$m or 100\,$\mu$m in the current implementation. The hydrogen flow speed is controlled by the heating of the hydrogen
generating a differential pressure in the reservoir and was set to a few millimeter per second.

The cryostat was attached on top of the ELFIE target chamber and could be moved in vertical direction with a motor. In this way, it could be moved upwards to clear the target chamber center and reference shots on different solid state targets could be performed without dismantling the hydrogen target setup. The ribbon was positioned and monitored in focal direction by imaging a shadowgram of the ribbon illuminated by a probe  beam at the second harmonic of the main laser ($\approx 527$\,nm) onto a camera. This beam was entering the target chamber from the side and crossing the main beam under an angle of 90 degree in the target chamber center (TCC). The energy spectra of the protons emitted at the target rear surface were recorded using radio-chromic film stacks (RCF) placed 5\,cm behind the target. A hole in the center of the stacks allowed for simultaneous measurement of the spectra with a Thomson parabola spectrometer aligned in target normal direction (see fig.\ref{setup}b)). Opening the chamber while the cryostat is still cold is not possible. In this case, water would condensate and freeze on the nozzle immediately. In order to remove the RCF stack after every shot, it was attached to a mechanical feedthrough and could be moved into a separately pumped transfer chamber. After closing a valve to the main chamber, the transfer chamber was vented and the stack removed.

To assess the efficiency of this novel target type to accelerate protons with respect to benchmarked solid-density target, 10\,$\mu$m thick gold and 6\,$\mu$m to 100\,$\mu$m thick plastic targets were also irradiated under the same laser conditions. 
In addition, the results are compared to a set of measurements utilizing metal targets with different thicknesses performed prior to this campaign at the ELFIE laser under comparable experimental conditions \cite{Romangani2015}.

\section{Measurements}

\begin{figure}
\centering
\includegraphics[width=.9 \textwidth]{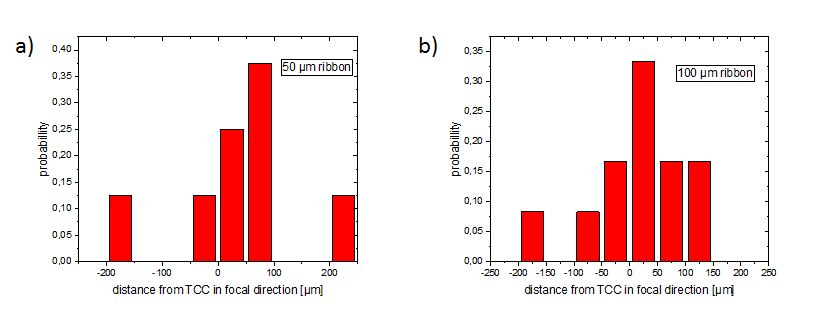}
\caption{Total position shift of the hydrogen ribbon relative to the target chamber center in laser direction. The position of the ribbon was monitored with the probe beam on every shot and the displacement with respect to TCC analyzed. Even though the 50\,$\mu$m thick ribbon (a) is subject to a stronger jitter than the 100\,$\mu$mm thick ribbon (b) on the time scale of a few seconds, the overall jitter magnitude of both target types is similar as it is dominated by the same slower movement of the entire ribbon on a larger spatial scale.}
\label{jitter}
\end{figure}

In a first series, the 50\,$\mu$m thick nozzle was installed onto the cryostat and several shots were taken. Prior to each shot, the target position was monitored on the side view camera and superimposed with the laser focus. The monitoring of the ribbon revealed a spatial movement along the focus direction of the main laser beam. In order to quantify this jitter, the position of the ribbon along the laser direction was recorded during a sequence of five seconds with a sample rate of 5\,Hz. For each frame in the sequence, the position of the ribbon was analyzed. Already on this short time scale, only in about 25\% of the measurements is the ribbon in a range of $\pm 25\,\mu$m around the focal spot. On a longer timescale of a minute an additional slower movement can be detected. This movement is most probably caused by bending of the target due to the large overall length of the ribbon of a few centimeters.  Since the timespan between moving the target into focus and launching a laser shot is in the range of a several minutes, the actual on-shot position was monitored in the sideview. Both displacements sum up to a total jitter that can even move the ribbon out of the field of view of the side view camera of 380\,$\mu$m as shown in figure \ref{jitter}. It turns out that many of the shots were likely out of focus. Therefore, in most of the laser shots no high energy protons could be detected. Nevertheless, proton energies higher then 10\,MeV were achieved 30\,\% of the shots. For a better comparison, only the two shots delivering the highest proton energy were taken into account for the further analysis.

In a second series of measurements the 100\,$\mu$m thick nozzle was used. In this case the fast jitter is substantially smaller. Within the short time window of a few seconds, the ribbon does not move out of the laser focus. Nevertheless, the slow large scale motion of the target still occurs, leading to laser shots with low proton energy. Thus, in case of the 100\,$\mu$m thick ribbon only the best performing shots were  analyzed as well.

\section{Results}

\begin{figure}
\centering
\includegraphics[width=1.\textwidth]{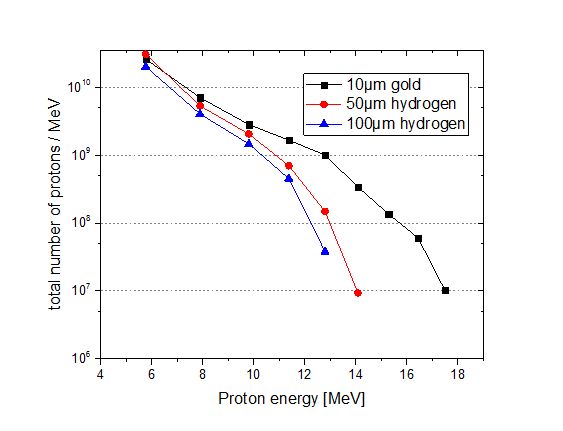}
\caption{Energy spectra for different hydrogen and gold targets recorded with RCF stacks: The two hydrogen ribbons give similar maximum proton energies of 12.8\,MeV and 14\,MeV for 100\,$\mu$m and 50\,$\mu$m, respectively.  The 10\,$\mu$m tick gold target accelerates protons to maximum energies of 18\,MeV.}
\label{spectrum}
\end{figure}

\begin{figure}
\centering
\includegraphics[width=.6 \textwidth]{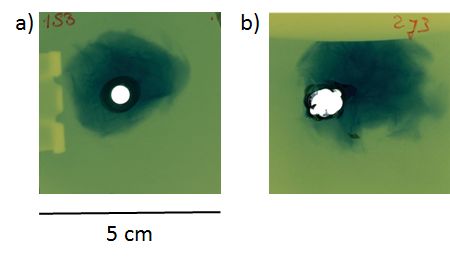}
\caption{Proton beam profile at 10\,MeV from a 10$\,\mu$m gold foil (a)) as well as a 50\,$\mu$m hydrogen ribbon (b)). Both beam profiles and divergences are similar. In case of the hydrogen ribbon, the shadow of the nozzle is visible in the upper part of the film.}
\label{profil}
\end{figure}

Figure\,\ref{spectrum} shows the proton energy spectra for the two best performing shots for the different hydrogen targets compared to a 10\,$\mu$m thick gold foil recorded with RCF stacks in forward direction. All three spectra show the expected exponential behavior typical for Target Normal Sheet Acceleration. The two hydrogen spectra have a similar shape and energy. The thinner ribbon (red circles) gives slightly higher maximum proton energies  of 14\,MeV as compared to 12.8\,MeV from the thicker ribbon (blue triangles). Furthermore, the proton number in the high energy range is increased by roughly $10\,\%$, which is within the uncertainty of the analysis. The gold foil (black squares) as reference leads to a maximum proton energy of 18\,MeV, 28\,$\%$ higher than the best performing hydrogen ribbon. Additionally, the overall proton number is increased. All three targets show similar proton beam profiles as can be seen in figure\,\ref{profil}. Both beams from the hydrogen ribbon have a divergence of roughly 30 degree full opening angle which is standard for this acceleration mechanism \cite{Bolton2014}. Since the ribbon was shot only a few millimeters below the cryostat, the shadow of the nozzle is visible in the proton beam on the upper part of the film.

\begin{figure}
\centering
\includegraphics[width=1. \textwidth]{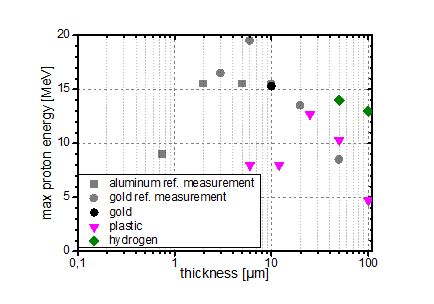}
\caption{Maximum energies for different target types: Both hydrogen ribbons show similar energies (green diamonds).
The gold and aluminum reference measurements (gray symbols) were taken in an earlier run \cite{Romangani2015} and verified with a 10\,$\mu$m gold target (black dot). The measurements for hydrogen are mean values from the best two shots each.}
\label{max_energies}
\end{figure}

A comparison of the maximum proton energies from hydrogen, metal and plastic targets is given in figure\,\ref{max_energies}. Each data point for the hydrogen ribbon marks the proton energy of the best performing shot for the specific target type.
Maximum proton energies from the hydrogen target, i.e. 12.8\,MeV and 14\,MeV from the 100\,$\mu$m and 50\,$\mu$m thick ribbon respectively, are marked by green diamonds.
Due to the low statistic it is not clear, if the slight increase in energy  with decreasing thickness marks a trend. Several energies for different metal targets are shown. Complementary to the 10\,$\mu$m thick gold foil (black triangle), maximum energies for aluminum and gold foils with thicknesses varying from 0.75\,$\mu$m up to 50\,$\mu$m from previous measurements at the same laser facility are plotted with gray symbols \cite{Romangani2015}. The metal targets show a maximum in proton energy of 19\,MeV at a foil thickness of 6\,$\mu$m and are in good agreement with the value determined from the 10$\,\mu$m gold target. For the hydrogen target, an influence of the laser contrast can not be seen for the used target dimensions. Decreasing the thickness could lead to an increase in proton energy. 

Additionally, the results of PET targets with an electron density between hydrogen and metal are shown. In this case, the target thickness was varied between 5\,$\mu$m and 100\,$\mu$m. The signal shows a clear maximum at 25\,$\mu$m. The maximum energy of 13\,MeV is even below the maximum energies from both the metal as well as the hydrogen target. According to TNSA scaling models, the proton energy is expected to increase with decreasing target thickness. 
However, this is usually limited by the laser temporal contrast: laser light reaching the target prior to the arrival of the main laser pulse can lead to significant pre-expansion of the target rear-surface plasma, followed by the degradation of accelerating gradients.
In a target thickness scan, thin targets are more sensitive to this effect than thick targets, therefore one usually observes an optimal target thickness, delivering highest proton energies\,\cite{Kaluza2004}. 
The optimal thickness of the heavy metal targets is around 6\,$\mu$m. In case of the lighter PET targets, the weakening of the accelerating field takes place already at 20\,$\mu$m and the proton energies decrease again. For the hydrogen, the target thickness was halved without seeing an significant effect in proton cut-off energies. Decreasing the target thickness further could lead to higher proton energies, which has to be tested in subsequent studies. Experiments with much thinner targets of 2\,$\mu$m from a cryogenic hydrogen jet with shorter laser pulses (30\,fs) already showed an equally good performance for hydrogen and metal \cite{Obst2017}.

\section{Conclusion}

In the presented experiments, we could show efficient proton acceleration from a hydrogen ribbon with an ultra-short pulse laser for the first time. Two different target thicknesses have been used, and maximum proton energies of 14\,MeV were reached with a 50\,$\mu$m thick target. In addition, the results were compared to gold and plastic targets. 

The generated proton beams have similar beam profiles and divergences like the commonly used metal or plastic targets. Although the proton energies are slightly lower than for the best performing gold foils, they are still within the same energy range. Since only two different target thicknesses could be tested, it was not possible to detect an optimal thickness for proton acceleration in the case of hydrogen. Decreasing the target thickness to values below 50\,$\mu$m could lead to higher proton energies. As seen before in an experiment at a long pulse laser system \cite{Margarone2016},  the number of protons with high energies exceeded the number reached with plastic foils. In addition, even the maximum energy was slightly higher.  In comparison to the best performing metal foils, the energies and particle numbers are slightly lower but still in the same range.  

The next steps towards routinely using the hydrogen ribbon as a proton source for applications have to solve three main challenges. Depending on the laser contrast, a thinner ribbon would be favorable for higher proton energies. This could be achieved with a thinner nozzle. Such a solution would require an even higher pressure in the hydrogen reservoir. Therefore it could be favorable to taper the ribbon after exiting the nozzle via controlled evaporation with an additional laser beam. In order to produce high energetic protons in every shot, it is necessary to improve the spatial stability of the target and suppress the long term movement. Shortening the ribbon with a laser beam tuned to an absorption wavelength of solid hydrogen could lead to the desired effect.

Furthermore, it showed that energy from the laser pulse is transported to the cryostat after every laser shot. This leads to an increase in temperature of the nozzle. Even though the temperature rise is below one Kelvin, in the case of the thin nozzle it is sufficient to melt the solid hydrogen and empty the whole reservoir after the shot. Cutting the ribbon between the nozzle and the interaction point could prevent the heat flow to reach the cryostat and inhibit the subsequent evaporation of the hydrogen reserve. 

The solid hydrogen target presented here has the potential for a high repetition rate target. Although only two different thicknesses could be used, proton beams obtained using the hydrogen ribbon were comparable to the ones from
the more commonly used metal foils. In addition, it meets the main criteria for targets used in applications with high repetition rates. Since the ribbon has a continuous flow, it is a  renewable target that does not require any target manipulation over several hours. Due to the production process of the solid hydrogen, the additional gas load into the chamber caused by the target is low. No additional debris that can harm the optics is produced, the only debris generated during the interaction becomes gas phase at ambient temperature and can be pumped down without accumulating on the optics.

This project has received funding from the European Union's Horizon 2020 research and innovation program under grant agreement no 654148 Laserlab-Europe and the German Government, Federal Ministry of education and research, grant no. 03ZIK445. This work was partly carried out within the frame of the project ANR-17-CE30-0026-Pinnacle from Agence Nationale de la Recherche, and partly within the framework of the EUROfusion Consortium and has received funding, through the ToIFE, from the European Union’s Horizon 2020 research and innovation programme under Grant Agreement No. 633053.  M. Gauthier acknowledges the support by the U.S. Department of Energy Office of Science, Fusion Energy Sciences under FWP 100182. The authors acknowledge valuable discussions with Thomas Kluge.

\nocite{*}
\bibliography{my_bib_h2_final}

\end{document}